\documentclass[nofootinbib,aps,11pt]{revtex4-1}
\usepackage{graphicx}
\usepackage{hyperref}
\usepackage{cancel}
\usepackage{amssymb}
\usepackage{textcomp}
\usepackage{amsmath}
\usepackage{bm}
\usepackage{times}
\usepackage{epsfig}
\usepackage{color}

\begin{document}
\title{\LARGE FIMP Dark Matter from Leptogenesis in Fast Expanding Universe}
\bigskip
\author{Zhi-Fang Chang}
\author{Zhao-Xuan Chen}
\author{Jia-Shu Xu}
\author{Zhi-Long Han}
\email{sps\_hanzl@ujn.edu.cn}
\affiliation{
School of Physics and Technology, University of Jinan, Jinan, Shandong 250022, China}
\date{\today}

\begin{abstract}
	Within the framework of canonical type-I seesaw, a feebly interacting massive particle (FIMP) $\chi$ is introduced as a dark matter candidate. The leptogenesis mechanism and dark matter relic density share a common origin via decays of Majorana  neutrinos $N$. Provided an additional species $\varphi$ whose energy density red-shifts as $\rho_{\varphi}\propto a^{-(4+n)}$, the Hubble expansion rate is larger than the standard scenario, i.e., the Universe expands faster. The consequences of such a fast expanding Universe on  leptogenesis as well as FIMP dark matter are investigated in detail. We demonstrate a significant impact on the final baryon asymmetry and dark matter abundance due to the existence of $\varphi$ for the strong washout scenario. While for the weak washout scenario, the effects of FEU are relatively small. We introduce scale factors $F_L$ and $F_\chi$ to describe the corresponding effects of FEU. A semi-analytical approach to derive the efficiency factors $\eta_L$ and $\eta_\chi$ in FEU is also discussed. The viable parameter space for success thermal leptogenesis and correct FIMP DM relic density is obtained for standard cosmology and FEU. Our results show that it is possible to distinguish different cosmology scenarios for strong washout cases.

\end{abstract}

\maketitle

\section{Introduction}

The observations of tiny neutrino mass, baryon asymmetry in the Universe, and dark matter (DM) have been  longstanding mysteries in particle physics.  The neutrino oscillations among different flavors are observed by atmospheric, solar, and reactor experiments \cite{Fukuda:1998mi,Ahmad:2002jz,An:2012eh}, which indicate that neutrino masses are sub-eV. Then, if neutrinos obtain masses via direct Yukawa interaction with standard Higgs doublet $y\bar{L}\tilde{H}\nu_R$, the Yukawa coupling $y\lesssim 10^{-12}$ is unnaturally small. Meanwhile, both precise measurements of Big Bang Nucleosynthesis (BBN) \cite{Iocco:2008va} and Cosmic Microwave Background (CMB) anisotropy \cite{Hu:2001bc} support the existence of an excess of baryonic matter over anti-baryonic matter. The measured baryon asymmetry normalized to the entropy density is $Y_B=(n_B-n_{\bar{B}})/s=8.72\pm0.04 \times 10^{-11}$ \cite{Aghanim:2018eyx}. Although the standard model (SM) has all the three ingredients of the Sakharov conditions, the generated baryon asymmetry is smaller than the observed value. As for dark matter, various astrophysical and cosmological observations such as galaxy rotation curves, the observation of bullet clusters, and the CMB anisotropy favor the existence of dark matter \cite{Bertone:2004pz}. The dark matter should be electrically neutral and stable or the decay lifetime longer than the age of the Universe. However, no SM particle can satisfy these characteristics with correct relic density. In summary, new physics beyond SM is required to interpret the origins of tiny neutrino mass, baryon asymmetry, and dark matter. 

One appealing way is to seek a common origin for these three issues, such as the $\nu$MSM \cite{Asaka:2005an,Asaka:2005pn,Abada:2018oly,Datta:2021elq}, the scotogenic model \cite{Ma:2006km,Ma:2006fn,Wang:2016lve,Hugle:2018qbw,Han:2019lux,Han:2019diw,Wang:2019byi,Guo:2020qin}, and the sterile neutrino portal model \cite{Escudero:2016tzx,Escudero:2016ksa,Falkowski:2017uya,Becker:2018rve,Liu:2020mxj,Coy:2021sse}. In this paper, we consider the third scenario. This model employs sterile Majorana neutrinos $N$ to generate tiny neutrino masses via the type-I seesaw mechanism \cite{Minkowski:1977sc,Mohapatra:1979ia}. The baryon asymmetry is generated via the thermal leptogenesis mechanism due to out-of-equilibrium CP-violation decays of $N$\cite{Fukugita:1986hr}. A dark sector with one scalar singlet $\phi$ and one Dirac fermion singlet $\chi$ is also introduced, which interact with SM particles via the heavy Majorana neutrinos $N$. Here, we consider $\chi$ is the FIMP DM candidate. For the scenario of WIMP and asymmetric DM see early studies in Ref.~ \cite{Escudero:2016ksa,Escudero:2016tzx,Falkowski:2011xh}. To make sure the stability of $\chi$, a $Z_2$ symmetry is imposed, which also helps to avoid the X-ray constraints \cite{Boyarsky:2018tvu}. Relic density of $\chi$ is then obtained due to $N\to \phi \chi$ decay via the freeze-in mechanism \cite{Falkowski:2017uya,Hall:2009bx}. The relevant Yukawa interactions and mass terms are
\begin{equation}
	-\mathcal{L}=y\bar{L}\tilde{H}N + \lambda \bar{\chi} \phi N + \frac{1}{2} \overline{N^C}m_N N+m_\chi \bar{\chi} \chi + \text{h.c.}.
\end{equation}

Noticeably, the above mentioned studies of the sterile neutrino portal model \cite{Escudero:2016ksa,Falkowski:2017uya,Liu:2020mxj} are based on the standard cosmology (SC). In this scenario, the Universe is radiation dominant after inflation until Big Bang Nucleosynthesis (BBN). However, since the cosmological history between inflation and BBN has no direct observational evidence, various non-standard cosmological models are allowed \cite{Barrow:1982ei,Kamionkowski:1990ni,Chung:1998rq,Salati:2002md,Profumo:2003hq,Pallis:2005hm,Arbey:2008kv,Arbey:2011gu,Gelmini:2013awa,Kane:2015qea,DEramo:2017gpl,DEramo:2017ecx,Bernal:2018kcw,Biswas:2018iny,Allahverdi:2019jsc,Chen:2019etb,Mahanta:2019sfo,Cosme:2020mck,Allahverdi:2020bys,Konar:2020vuu,Barman:2021ifu}. In the modified cosmological scenario, the expansion rate of the Universe, i.e., the Hubble parameter, deviates from the standard scenario. For instance, introducing a scalar filed $\varphi$, whose energy density red-shifts with the scale factor a as $\rho_{\varphi}\propto a^{-(4+n)}$, can lead to a fast expanding Universe (FEU) when $n>0$ \cite{DEramo:2017gpl}. We note that the period of FIMP DM generation from leptogenesis is right between inflation and BBN. Therefore, modification of the cosmology history will affect the evolution of lepton asymmetry \cite{Chen:2019etb} and DM abundance \cite{DEramo:2017ecx}. Based on several benchmark points, it has been shown that the lepton asymmetry $Y_L$ and DM abundance $Y_\chi$ can be changed by several orders of magnitudes \cite{Chen:2019etb,DEramo:2017ecx}. This indicates that the viable parameter space for success leptogenesis and FIMP DM in modified cosmology would be quite different from  the standard scenario, which can then be used to distinguish different cosmology models.  In this work, we endeavor to comprehensively investigate the viable parameter space of FIMP DM from leptogenesis in FEU.

The paper is organized as follows. In Sec. \ref{Sec:FEU}, the framework of the fast expanding Universe is briefly reviewed. Evolution of FIMP DM and leptogenesis in the FEU for some benchmark points is presented in Sec. \ref{Sec:DM}. Then, we perform a random scan over the parameter space in Sec. \ref{Sec:DS}. The viable parameter space for success thermal leptogenesis and correct FIMP DM relic density is obtained for standard cosmology and FEU. Conclusions are in Sec. \ref{Sec:CL}.

\section{A Fast Expanding Universe}\label{Sec:FEU}
In the standard cosmology, radiation is the dominant component before BBN, whose energy density is given by
\begin{align}\label{Eqn:rho}
\rho_{r}(T)=\frac{\pi^2}{30}g_*(T)T^4,
\end{align}
where $T$ is the temperature of the Universe and $g_{*}(T)$ is the effective energy degrees of freedom. The Hubble parameter, which describes the expanding rate of the Universe, is related to the radiation as
\begin{align}
H_r(T)=\sqrt{\frac{8\pi G\rho_{r}(T)}{3}}=1.66\sqrt{g_*(T)}\frac{T^2}{M_p},
\end{align}
with the Planck mass $M_p=1.22\times10^{19}$ GeV.

In a fast expanding Universe, an additional scalar component $\varphi$ coexists with the radiation. We assume the energy density of $\varphi$ scales as 
\begin{align}
\rho_{\varphi}\sim a^{-(4+n)},n>0.
\end{align}
It is convenient to express $\rho_{\varphi}$ as a function of the temperature $T$. This can be achieved by assuming entropy conservation in a coming volume $S=sa^3=$ constant, where the entropy density of the universe is parameterized as
\begin{align}
s(T)=\frac{2\pi^2}{45}g_{*s}(T)T^3,
\end{align}
with  $g_{*s}(T)$ being the effective entropy degrees of freedom. This makes sure $g_{*s}(T)T^3 a^3=$ constant, and the energy density of $\varphi$ is derived as
\begin{align}
\rho_\varphi(T)=\rho_\varphi(T_r)\left(\frac{g_{*s}(T)}{g_{*s}(T_r)}\right)^{\frac{4+n}{3}}\left(\frac{T}{T_r}\right)^{4+n}.
\end{align}
Here, $T_r$ is defined as the temperature when the two components have equal energy density, i.e., $\rho_{\varphi}(T_r)=\rho_{r}(T_r)$. Using the relation $\rho_r(T)\propto g_*(T)T^4$ in Eqn. \eqref{Eqn:rho}, we further obtain
\begin{equation}
	\rho_\varphi(T)=\rho_r(T)\frac{g_*(T_r)}{g_*(T)}\left(\frac{g_{*s}(T)}{g_{*s}(T_r)}\right)^{\frac{4+n}{3}}\left(\frac{T}{T_r}\right)^{n}.
\end{equation}
Therefore, the total energy density can be expressed as
\begin{align}
\rho(T)=\rho_{r}(T)+\rho_{\varphi}(T)=\rho_{r}(T)\left[1+\frac{g_*(T_r)}{g_*(T)}\left(\frac{g_{*s}(T)}{g_{*s}(T_r)}\right)^{\frac{4+n}{3}}\left(\frac{T}{T_r}\right)^n\right].
\end{align}
For temperature $T>T_r$, $\rho_{\varphi}(T)>\rho_{r}(T)$, the Universe is dominant by the component $\varphi$.
Considering the typical temperature for success leptogenesis $T\sim M_N\gtrsim 10^9$ GeV in the canonical type-I seesaw, we can take $g_*(T)=g_{*s}(T)=106.75$ as a constant. The total energy density is then simplified as
\begin{align}
\rho(T)=\rho_{r}(T)\left[1+\left(\frac{T}{T_r}\right)^n\right].
\end{align}
Finally, the Hubble parameter in a fast expanding Universe  is  modified by
\begin{align}\label{eq:Hp}
H(T)=1.66\sqrt{g_*}\frac{T^2}{M_p}\left[1+\left(\frac{T}{T_r}\right)^n\right]^\frac{1}{2}=H_r(T)\left[1+\left(\frac{T}{T_r}\right)^n\right]^\frac{1}{2}.
\end{align}
It is clear that for $T\gg T_r$, $H(T)\simeq H_r(T)(T/T_r)^{n/2}$, the Universe is expanding faster than the radiation dominant scenario. Meanwhile, for $T\ll T_r$, $H(T)\simeq H_r(T)$, the Universe recovers the standard one.  It should be noticed that for the scenario with $n=0$, we do not obtain the standard cosmology but a modified one still with a new component that red-shifts like radiation. 

In order not to conflict with observations from BBN, a lower limit on $T_r$ \cite{DEramo:2017gpl}
\begin{equation}
	T_r\geq (15.4)^{1/n} ~\text{MeV},
\end{equation}
must be satisfied. In this paper, we consider $T_r$ around the scale of canonical thermal leptogenesis ($\gtrsim 10^{9}$~GeV), thus the above BBN limit is always satisfied.

\section{FIMP DM from Leptogenesis in FEU}\label{Sec:DM}

Before study the phenomenology of FIMP DM from leptogenesis in FEU, we review the theoretical framework in the standard cosmology. The CP asymmetry is generated by out-of-equilibrium CP-violating decays of Majorana neutrino. Neglecting the flavor effect \cite{Nardi:2006fx} and considering a hierarchy mass structure of heavy Majorana neutrinos  $M_1\ll M_2, M_3$, the CP asymmetry is
\begin{equation}
	\varepsilon =-\frac{3}{16 \pi (y^{\dagger} y)_{11}} \sum_{j=2,3} \operatorname{Im}\left[\left(y^{\dagger} y\right)_{1 j}^{2}\right] \frac{M_{1}}{M_{j}}.
\end{equation}
The Yukawa matrix $y$ can be related with light neutrino oscillation parameters by using the Casas-Ibarra parametrization \cite{Casas:2001sr,Ibarra:2003up}
\begin{equation}\label{eq:CI}
y=\frac{\sqrt{2}}{v}U_{\text{PMNS}}\hat{m}_\nu^{1/2} R (\hat{m}_{N})^{1/2},
\end{equation}
where $U_{\text{PMNS}}$ is the neutrino mixing matrix, $R$ is an orthogonal matrix, and $\hat{m}_\nu=\mbox{diag}(m_1,m_2,m_3)$ [$\hat{m}_N=\mbox{diag}(M_1,M_2,M_3)$] is the diagonal light (heavy) neutrino mass matrix. Known as the Davidson-Ibarra bound \cite{Davidson:2002qv}, an upper limit on $\varepsilon$ can be derived
\begin{equation}\label{eq:eps1}
|\varepsilon|\lesssim \frac{3}{16\pi}\frac{M_1 m_3}{v}.
\end{equation}

The evolution of abundance $Y_{N_1},Y_\chi$ and lepton asymmetry $Y_{L}$ in the standard cosmology is described by the Boltzmann equations
\begin{eqnarray}
\frac{dY_{N_1}}{dz} &=& - D_r (Y_{N_1}-Y_{N_1}^{eq}),\\
\frac{dY_{L}}{dz} &=&- \varepsilon D_r (Y_{N_1}-Y_{N_1}^{eq})- W_r Y_{L}, \\
\frac{dY_{\chi}}{dz} &=&  D_r ~Y_{N_1} \text{BR}_\chi,
\end{eqnarray}
where $z=M_1/T$. $\text{BR}_\chi$ is the branching ratio of $N_1\to \phi \chi$. The decay and washout terms are
\begin{equation}
	D_r(z)=K z \frac{\mathcal{K}_1(z)}{\mathcal{K}_2(z)}, W_r(z)= \frac{1}{4} K z^3 \mathcal{K}_1(z),
\end{equation}
where $\mathcal{K}_{1,2}(z)$ are the modified Bessel function. To quantify the washout effect, the decay parameter $K$ is defined as
\begin{equation}
K = \frac{\Gamma_1}{H_r(z=1)},
\end{equation}
with the decay width of $N_1$ being $\Gamma_1=(y^\dagger y)_{11} M_1/(8\pi)$.

Latter, the lepton asymmetry is converted into the baryon asymmetry via the sphaleron processes with the relation \cite{Davidson:2008bu}
\begin{align}
Y_B=\frac{28}{79}Y_L.
\end{align} 

In a fast expanding Universe, the Hubble parameter is modified as in Eqn. ~\eqref{eq:Hp}. Then, the corresponding Boltzmann equations have the same form as the standard cosmology
\begin{eqnarray}
\frac{dY_{N_1}}{dz} &=& - D  (Y_{N_1}-Y_{N_1}^{eq}),\\
\frac{dY_{L}}{dz} &=&- \varepsilon D  (Y_{N_1}-Y_{N_1}^{eq})- W  Y_{L} , \\
\frac{dY_{\chi}}{dz} &=&  D  ~Y_{N_1} \text{BR}_\chi,
\end{eqnarray}
but the decay and washout terms are modified as
\begin{equation}
D(z)=K z \frac{\mathcal{K}_1(z)}{\mathcal{K}_2(z)} \left[1+\left(\frac{z_r}{z}\right)^{n}\right]^{-1 / 2}, W(z)= \frac{1}{4} K z^3 \mathcal{K}_1(z) \left[1+\left(\frac{z_r}{z}\right)^{n}\right]^{-1 / 2},
\end{equation}
where the parameter $z_r$ is defined as $z_r\equiv M_1/T_r$. It is clear that the effect of the new component $\varphi$ is described by the two parameter $z_r$ and $n$. In the scenario with $z_r\ll 1$, i.e., $M_1\ll T_r$, radiation is the dominant component at the scale of leptogenesis $T\sim M_1$, which resembles leptogenesis in the standard cosmology. Therefore, we consider the scenario with $z_r\gtrsim 1$, i.e., $M_1\gtrsim T_r$, where large observational effect on leptogenesis is expected.

\begin{figure}
	\centering
	\includegraphics[width=0.95\textwidth]{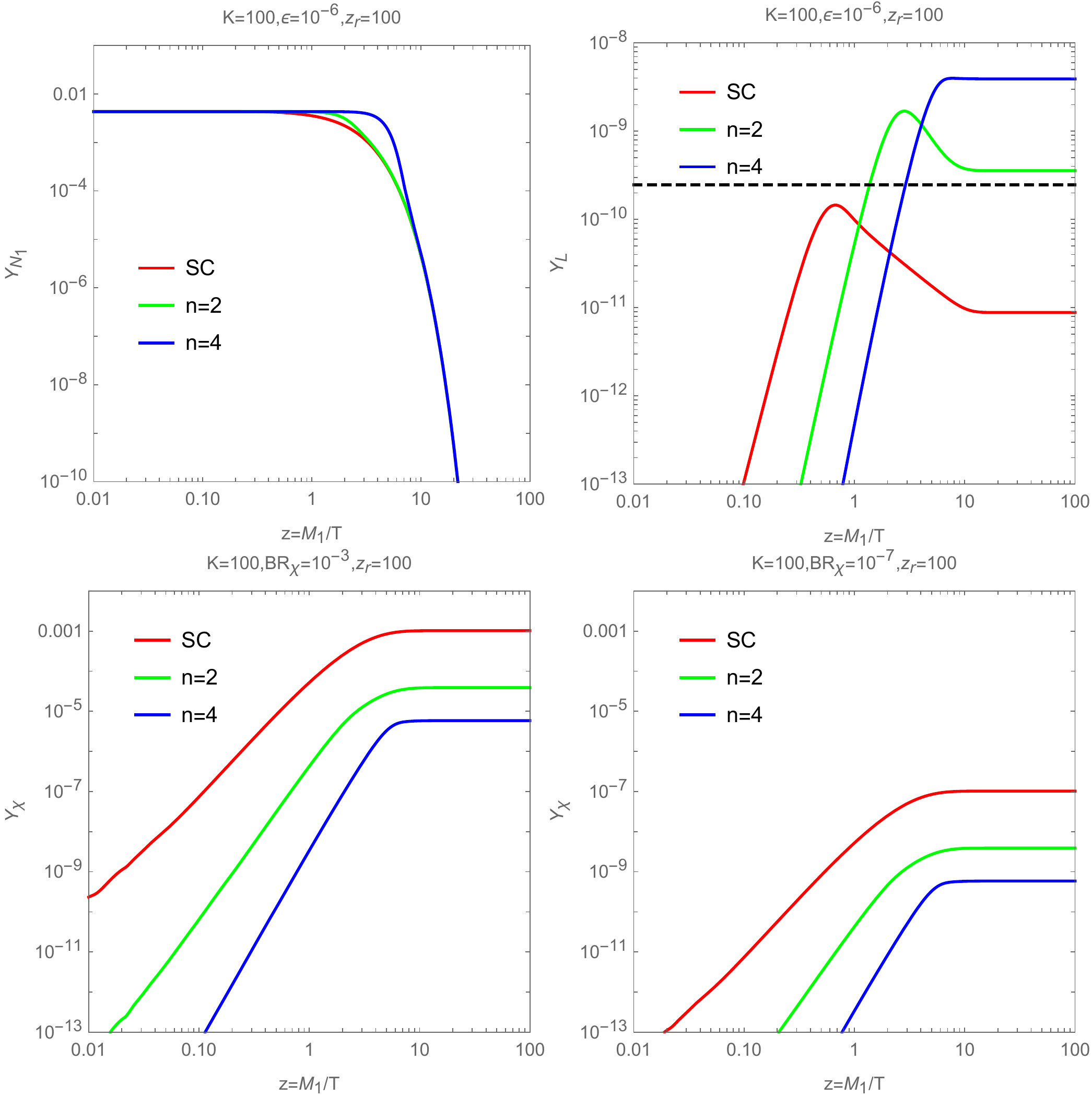}
	\caption{Evolution of abundance $Y_{N_1}$, $Y_L$, and $Y_\chi$ for strong washout scenario with $K=100$. 
		\label{FIG:YNS}} 	
\end{figure}

Now, let's explicitly illustrate the effect of the fast expanding Universe for some benchmark points. We assume that the initial abundance of Majorana neutrinos $N$ are in thermal equilibrium $Y_{N_1}^\text{in}=Y_{N_1}^\text{eq}$, and the initial lepton asymmetry and DM abundance are vanishing $Y_L^\text{in}=0,Y_\chi^\text{in}=0$. In Fig. \ref{FIG:YNS}, the effects of parameter $n$ on the evolution of abundance  $Y_{N_1}$, $Y_L$, and $Y_\chi$ for strong washout scenarios are presented. During the calculation, we have fixed $K=100$, $\varepsilon=10^{-6}$, and $z_r=100$. Meanwhile, two different branching ratios of FIMP DM, i.e., $\text{BR}_\chi=10^{-3},10^{-7}$, are shown. Comparing with the standard cosmology, a larger $n$ will push $N_1$ out-of-equilibrium more efficiently, which thus leads to an enhancement of the lepton asymmetry. In the up-right panel of Fig. \ref{FIG:YNS}, it shows that the evolution of $Y_L$ is dramatically modified in FEU, where the horizontal line denotes the required $Y_L$ to generate the observed $Y_B$. Qualitatively speaking,  for standard cosmology, the final lepton asymmetry of this benchmark scenario is far below the required value. But for FEU with $n=2$, the washout effect is reduced appropriately, which results in proper lepton asymmetry. As for FEU with $n=4$, the expansion of the Universe is too fast that the evolution of $Y_L$ behaves as a weak washout scenario. Hence, the final lepton asymmetry is larger than the required value. Opposite to $Y_L$, the FIMP DM abundance $Y_\chi$ decreases as the parameter $n$ increases. Therefore, the production of DM $\chi$ from $N$ decay becomes less efficient when the Universe expands faster. As for different branching ratios $\text{BR}_\chi$, the final abundances $Y_\chi$ are proportional to $\text{BR}_\chi$ with the same dependence on the parameter $n$. 

\begin{figure}
	\centering
	\includegraphics[width=0.95\textwidth]{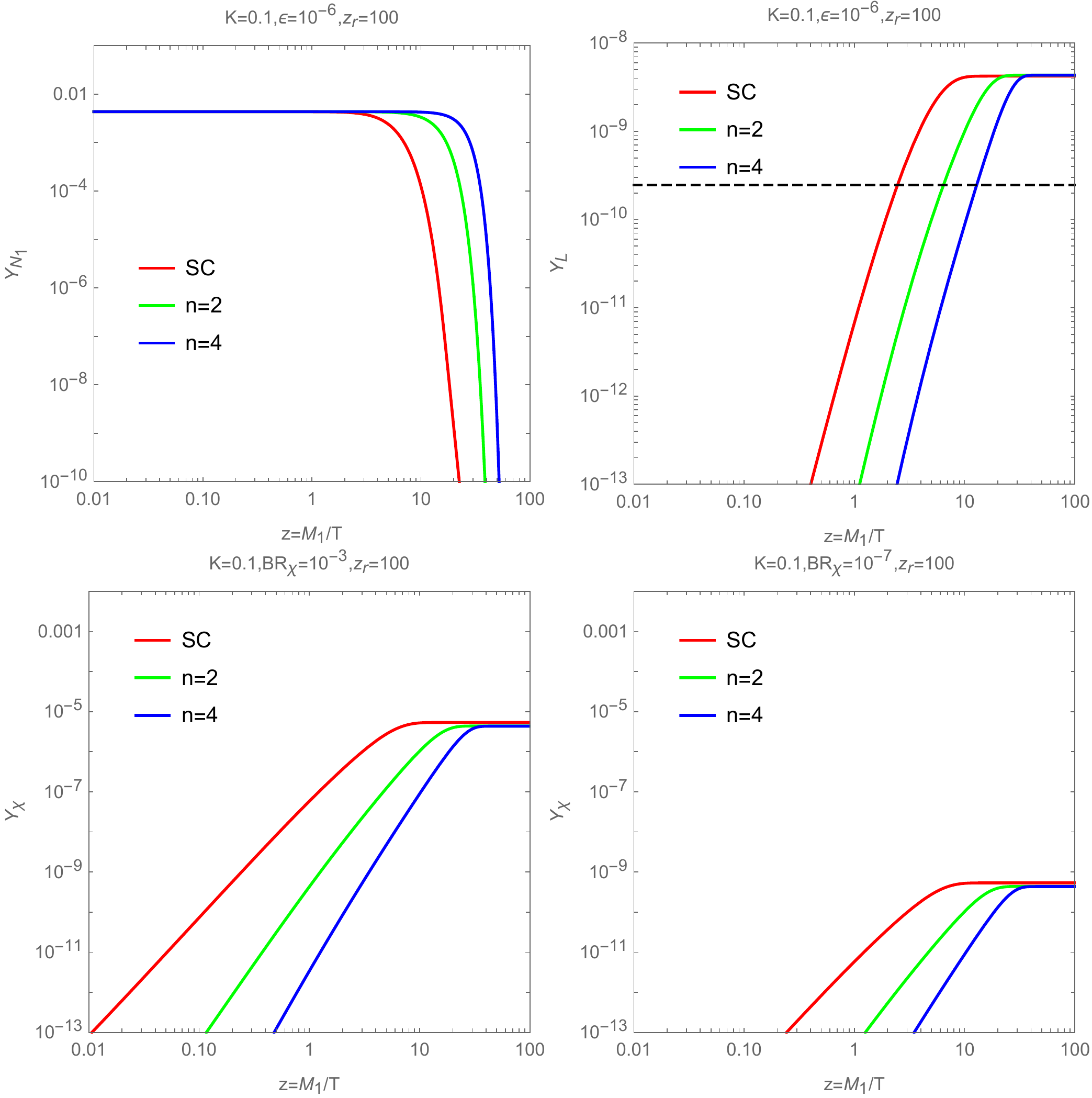}
	\caption{Evolution of abundance $Y_{N_1}$, $Y_L$, and $Y_\chi$ for weak washout scenario with $K=0.1$. 
		\label{FIG:YNW}} 	
\end{figure}

In Fig.~\ref{FIG:YNW}, the effects of parameter $n$ for weak washout scenario with $K=0.1$ are shown. Comparing with the standard cosmology, the decay of $N_1$ is delayed due to faster expansion of the Universe with increasing $n$. Therefore, the production of lepton asymmetry $Y_\chi$ and DM abundance $Y_\chi$ are also postponed. But the final lepton asymmetry and DM abundance are of the same order for different $n$.  Our numerical results show that the relative differences between SC and FEU for lepton asymmetry $Y_L$ and DM abundance $Y_\chi$ are less than 2.6\% and 19\% respectively. This indicates that for the weak washout scenario, a fast expanding Universe does not have a great impact on the final $Y_L$ and $Y_\chi$ when varying parameter $n$ \footnote{Note that if we consider the initial $Y_{N_1}^\text{in}=0$, the final $Y_L$ and $Y_\chi$ will also be greatly modified for different values of $n$ in the weak washout scenario.} . 

\begin{figure} 
	\centering
	\includegraphics[width=0.95\textwidth]{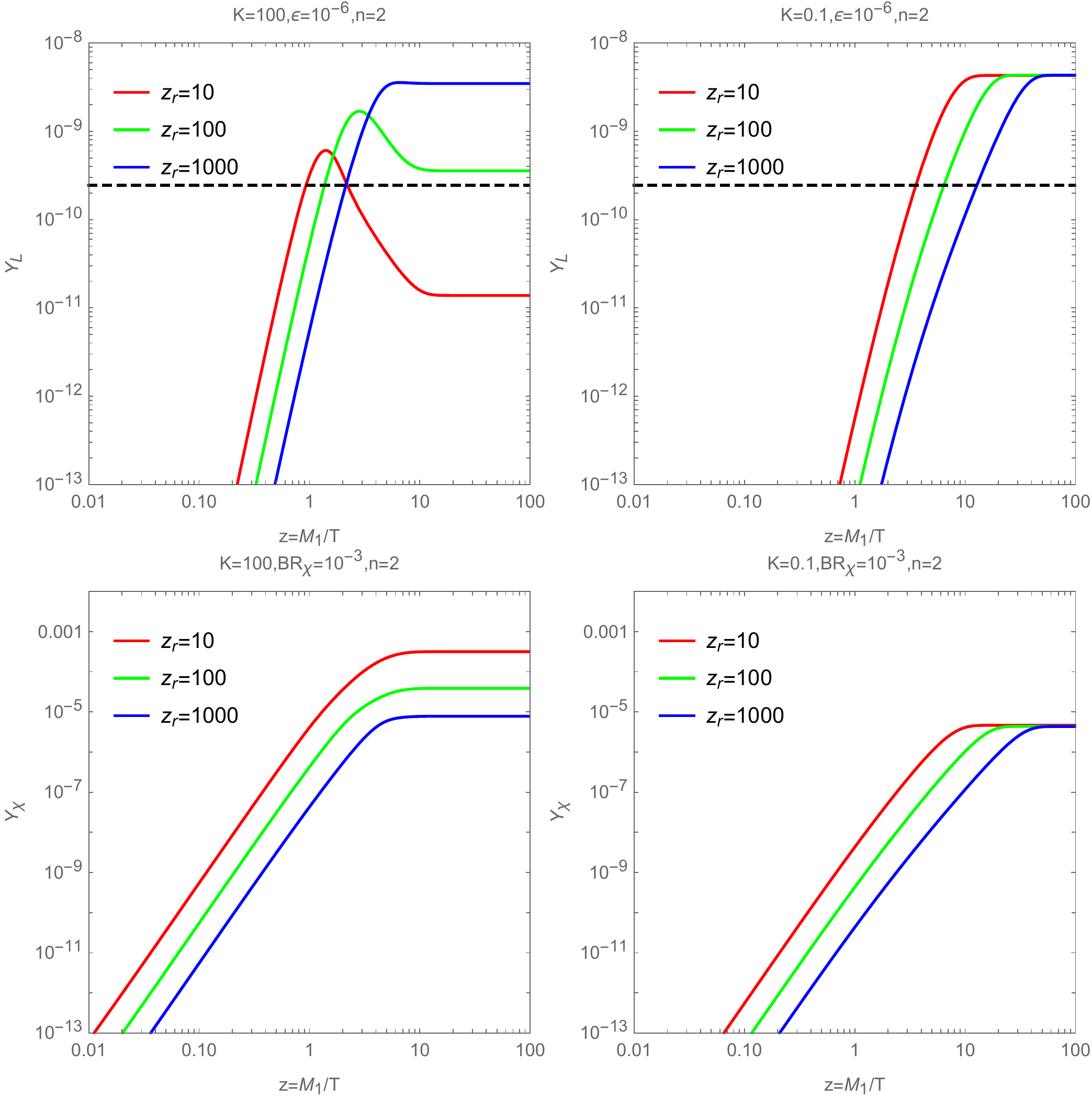}
	\caption{Effects of parameter $z_r$ on the evolution of $Y_L$ and $Y_\chi$ for strong (left) and weak (right) washout scenarios.
		\label{FIG:YLTr}} 	
\end{figure}

Effects of another parameter $z_r$ on the evolution of lepton asymmetry $Y_L$ and DM abundance $Y_\chi$ are shown in Fig.~\ref{FIG:YLTr} for the strong ($K=100$) and weak ($K=0.1$) washout scenario. The other relevant parameters are fixed as $n=2$, $\varepsilon=10^{-6}$ and $\text{BR}_\chi=10^{-3}$. In the strong washout scenario, modifications of $z_r$ will have a great impact on the evolution of $Y_L$ and $Y_\chi$. For instance, the evolution of $Y_L$ for $z_r=10$ behaves much similar to the standard evolution with the final $Y_L$ much lower than the required value. Meanwhile, for $z_r=1000$, the washout effect is disappearing with the final $Y_L$ enhanced by more than two orders of magnitudes. The final DM abundance $Y_\chi$ becomes much smaller as $z_r$ goes larger. Same as the impact of parameter $n$, the final $Y_L$ and $Y_\chi$ are slightly modified for different choices of $z_r$ in the weak washout scenario. By directly comparing with the results for strong and weak washout scenarios, it is clear that the decay parameter $K$ would lead to different evolution behavior of lepton asymmetry $Y_L$ and DM abundance $Y_\chi$. 

Based on the above discussion, we can conclude that the final lepton asymmetry $Y_L$ (DM abundance $Y_\chi$) increases (decreases) as the parameter $n$ or $z_r$ increase for the strong washout scenario. As for the weak washout case, modifications of parameter $n$ or $z_r$ delay the production of $Y_L$ and $Y_\chi$, but will not greatly affect the final $Y_L$ and $Y_\chi$. To quantitatively describe the effect of the fast expending Universe on the lepton asymmetry $Y_L$ and DM abundance $Y_\chi$, we define the following scale factors
\begin{equation}\label{eq:SF}
	F_L=\frac{Y_L^\text{FEU}(\infty)}{Y_L^\text{SC}(\infty)},\quad F_\chi=\frac{Y_\chi^\text{FEU}(\infty)}{Y_\chi^\text{SC}(\infty)}.
\end{equation}

\begin{figure} 
	\centering
	\includegraphics[width=0.95\textwidth]{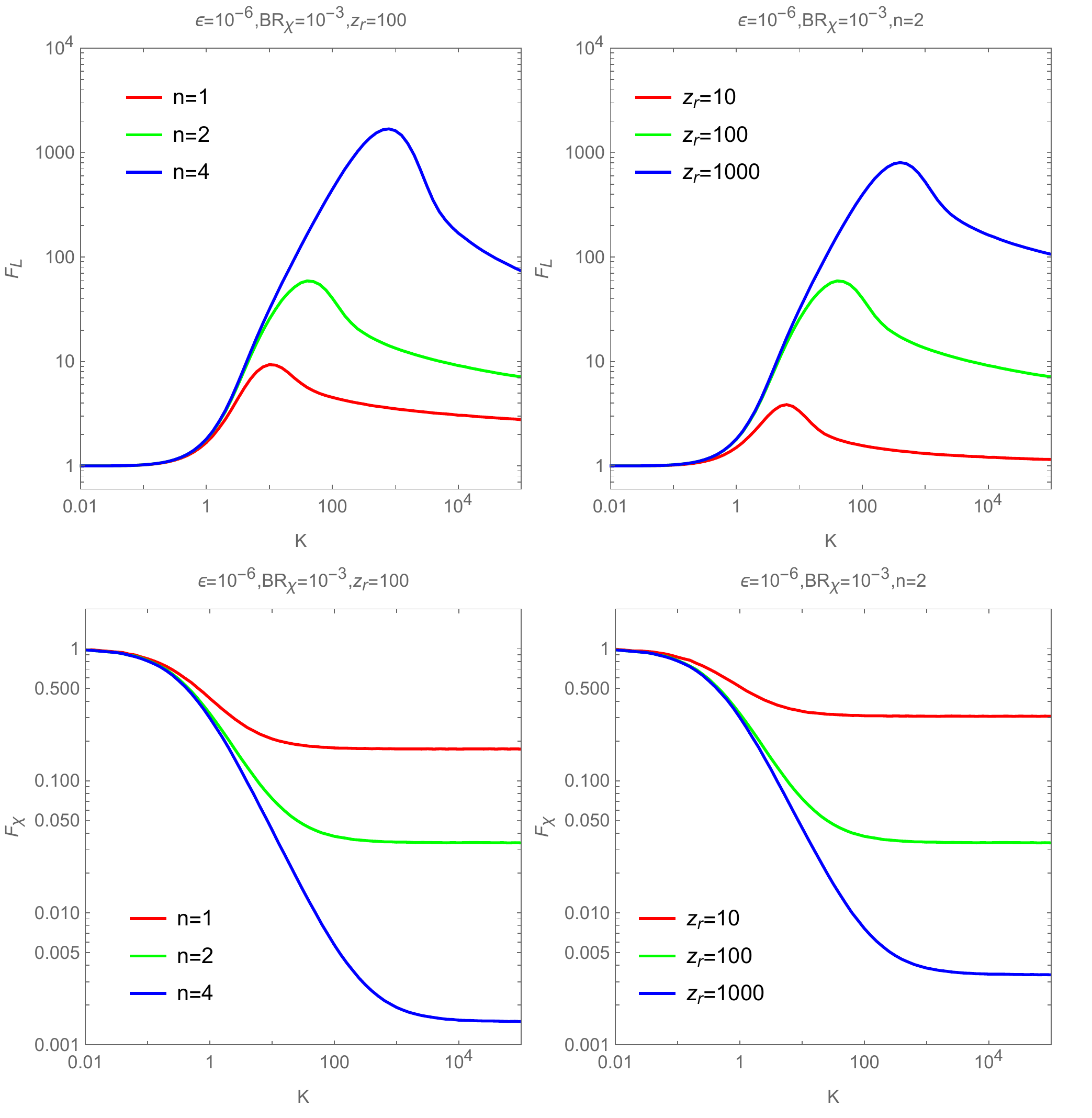}
	\caption{Effects of parameters $n$ (left) and $z_r$ (right) on the scale factors $F_L$ and $F_\chi$ as a function of $K$.
		\label{FIG:SF}} 	
\end{figure}

The numerical results for $F_L$ and $F_\chi$ as a function of $K$ are shown in Fig.~\ref{FIG:SF}. We have fixed $z_r=100$ when studying the effects of $n$. Meanwhile, $n=2$ is chosen to illustrate the effects of $z_r$. It is clear that $F_L\gtrsim 1$ and $F_\chi\lesssim 1$ for the weak washout scenario $K\lesssim 1$. For  intermediate case $1\lesssim K\lesssim 10$, $F_L(n=2)$ exactly equals to  $F_L(n=4)$, which means that to observe differences between these two choices of $n$, a strong washout scenario with $K\gtrsim 10$ is required. A peak value of $F_L$ is also observed. Specifically speaking, the peak values of $F_L$ are approximately 10 at $K\sim10$, 60 at $K\sim40$, and 1700 at $K\sim 780$ for $n=1,2,4$ respectively. When the parameter $K$ is large enough, the scale factor $F_L$ then decreases as $K$ increases. Similar effects on $F_L$ are also observed when varying the parameter $z_r$. For instance, $F_L$ reaches a peak of 800 at $K\sim 400$ for $z_r=1000$. As for the scale factor $F_\chi$, it first decreases as $K$ increases. Then $F_\chi$ approaches to a constant for large enough $K$, e.g., $F_\chi\sim 0.17,0.034,0.0015$ for $n=1,2,4$ respectively. Varying $z_r$ leads to a similar effect on $F_\chi$. For large $z_r$ and $K$, Ref.~\cite{DEramo:2017ecx} has also derived an analytical expression for $F_\chi$,
\begin{equation}
	F_\chi=\frac{8}{3\pi}\left(\frac{2}{z_r}\right)^{n/2}\Gamma\left(\frac{6+n}{4}\right)\Gamma\left(\frac{10+n}{4}\right).
\end{equation}

\begin{figure} 
	\centering
	\includegraphics[width=0.95\textwidth]{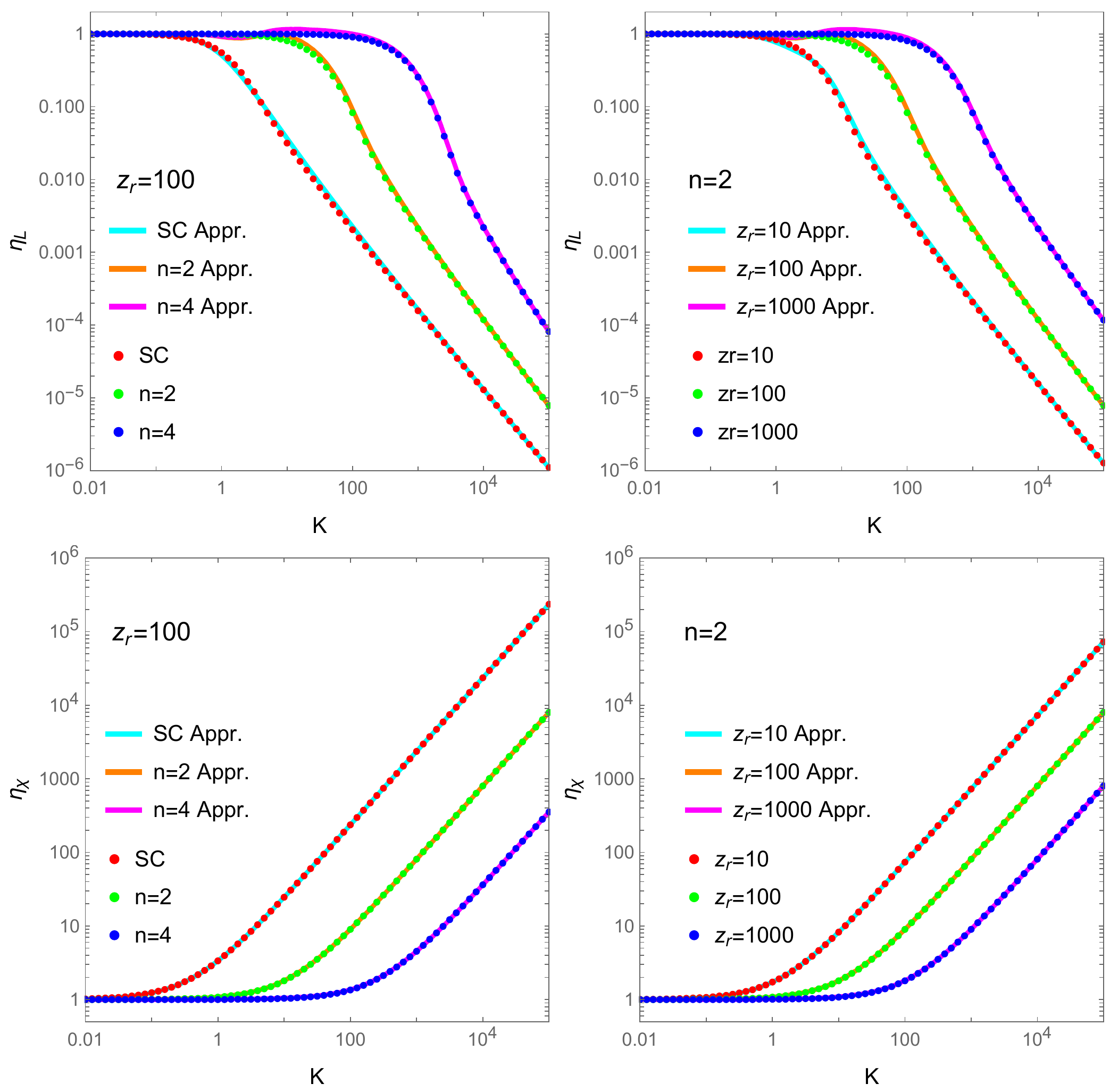}
	\caption{Effects of parameters $n$ (left) and $z_r$ (right) on the efficiency factors $\eta_L$ and $\eta_\chi$ as a function of $K$. The data points correspond to the numerical results, while the solid lines correspond to the approximate expressions in Eqn.~\eqref{eq:EF1} and \eqref{eq:EF2}.
		\label{FIG:EF}} 	
\end{figure}

It is well known that the final lepton asymmetry $Y_L(\infty)$ and DM abundance $Y_\chi(\infty)$ can be conveniently parametrized as \cite{Buchmuller:2004nz,Falkowski:2017uya}
\begin{equation}
	Y_L(\infty)=\epsilon\,\eta_L Y_{N_1}^\text{eq}(0),~Y_\chi(\infty)=\text{BR}_\chi\,\eta_\chi Y_{N_1}^\text{eq}(0),
\end{equation}
where $\eta_L,\eta_\chi$ are called as efficiency factor. In the standard cosmology, these two factors can be approximately expressed as
\begin{equation}\label{eq:EF1}
	\eta_L^\text{SC}\simeq\frac{1}{1+1.3K \ln (1+K)^{0.8}},~\eta_\chi^\text{SC}\simeq1+\frac{3\pi}{4}K.
\end{equation}
It is obvious that these efficiency factors only depend on the parameter $K$. Therefore, from the definition of scale factors $F_L$ and $F_\chi$ in  Eqn.~\eqref{eq:SF}, we are also aware that $F_L$ and $F_\chi$ are functions of parameters $K,n$ and $z_r$, and are independent of parameters $\epsilon$ and BR$_\chi$. Using the scale factors, we can easily derive the semi-analytical expressions of efficiency factors in FEU as
\begin{equation}\label{eq:EF2}
	\eta_L^\text{FEU}=\eta_L^\text{SC} F_L^\text{FEU},~\eta_\chi^\text{FEU}=\eta_\chi^\text{SC} F_\chi^\text{FEU}.
\end{equation}

The numerical and approximate results are both shown in Fig.~\ref{FIG:EF}. As clearly shown, the approximate expressions in Eqn.~\eqref{eq:EF1} and \eqref{eq:EF2} can well fit the numerical results for all cases. Basically speaking, the efficiency factor $\eta_L$ ($\eta_\chi$) could be enhanced (suppressed) by several orders of magnitude due to FEU for the strong washout scenario. Meanwhile for weak washout case $K\lesssim1$, we always have $\eta_L\sim1$ and $\eta_\chi\sim 1$. Inversely, if we named the effective weak washout scenario with the condition $\eta_L\lesssim 1$, then $K\lesssim30(550)$ for $n=2(4), z_r=100$ or $K\lesssim3(270)$ for $n=2,z_r=10(1000)$ will satisfy the requirement.

\section{Discussion}\label{Sec:DS}

\begin{figure} 
	\centering
	\includegraphics[width=0.95\textwidth]{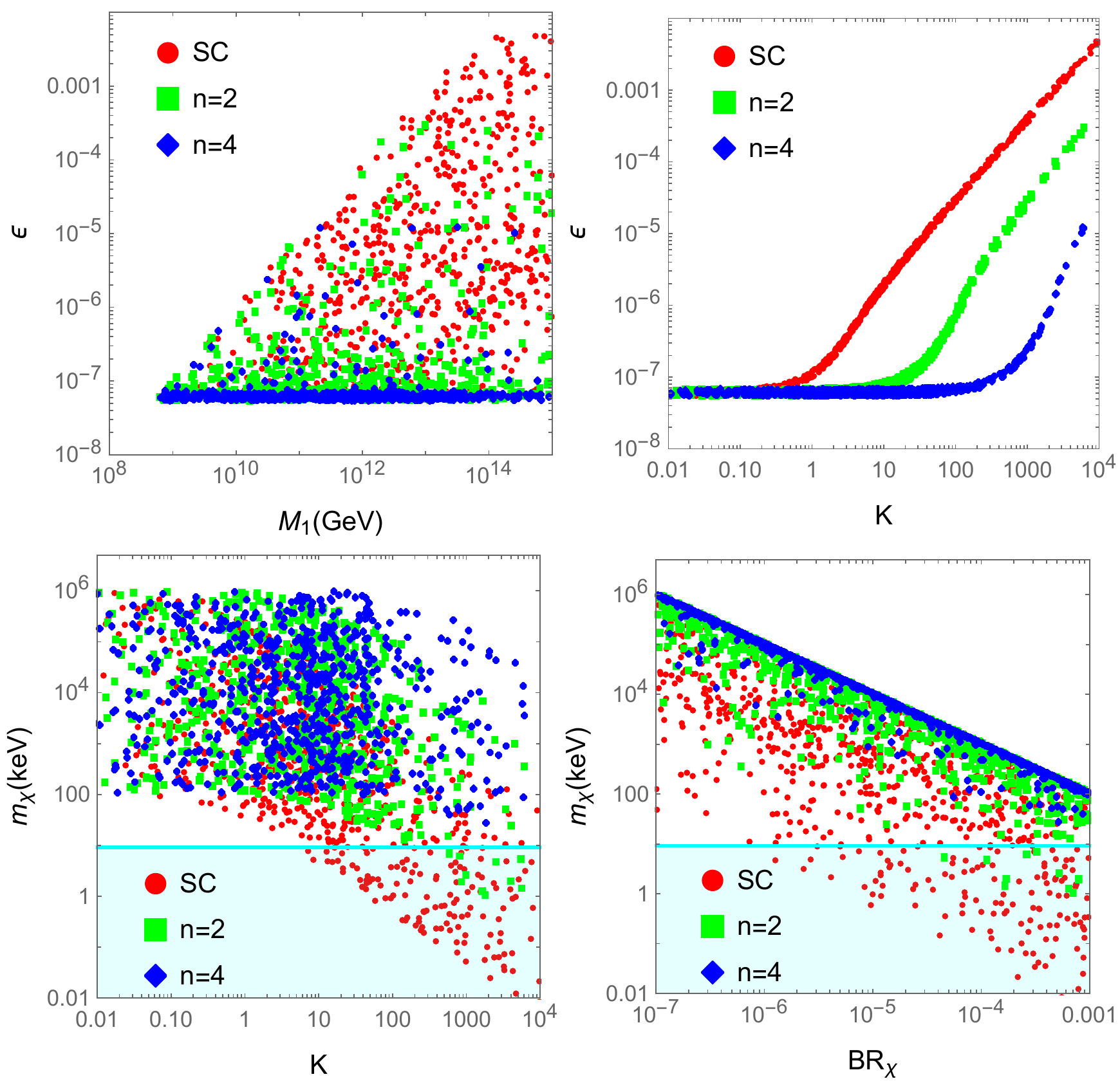}
	\caption{Scanned results for SC and FEU with n=2, 4. For FEU, we take $z_r=100$ to illustrate.
		\label{FIG:Scan1}} 	
\end{figure}

In the above section, we have shown that the lepton asymmetry $Y_L$ and DM abundance $Y_\chi$ would be greatly modified in FEU, especially for the strong washout scenario. Since the Planck collaboration has already given the precise measurements \cite{Aghanim:2018eyx}
\begin{equation}
	Y_B=(8.72\pm 0.04)\times 10^{-11},~\Omega_\chi h^2=0.120\pm 0.001.
\end{equation}
We further explore the parameter space that can predict correct baryon asymmetry and DM relic density.
A random scan over the following parameter space is performed
\begin{eqnarray}
	m_1\in [10^{-5},0.03]~\text{eV},~M_1\in [10^8,10^{15}]~\text{GeV},\\
	\text{Re}(\omega_{ij}),\text{Im}(\omega_{ij})\in[10^{-5},\pi],\text{BR}_\chi\in[10^{-7},10^{-3}],
\end{eqnarray}
where $\omega_{ij}=\omega_{12},\omega_{13},\omega_{23}$ are the three complex angles of the orthogonal matrix $R$ in Eqn.~\eqref{eq:CI}. Explicit expressions of $R$ can be found in Ref.~\cite{Liu:2020mxj}.
Here, we have assumed a normal hierarchy of neutrino mass structure. The upper limit on lightest neutrino mass $m_1<0.03$ eV is obtained from the requirement of $\sum m_i<0.12$ eV \cite{Aghanim:2018eyx}. During the scan, $M_2/M_1=M_3/M_2=10$ is assumed. As mentioned in Ref.~\cite{Hugle:2018qbw}, unflavored leptogenesis is insensitive to the neutrino mixing matrix. Therefore, the neutrino oscillation parameters are fixed to the global best fit values provided in Ref.~\cite{deSalas:2020pgw}. The two additional Majorana phases are also fixed to be zero.

\begin{figure} 
	\centering
	\includegraphics[width=0.95\textwidth]{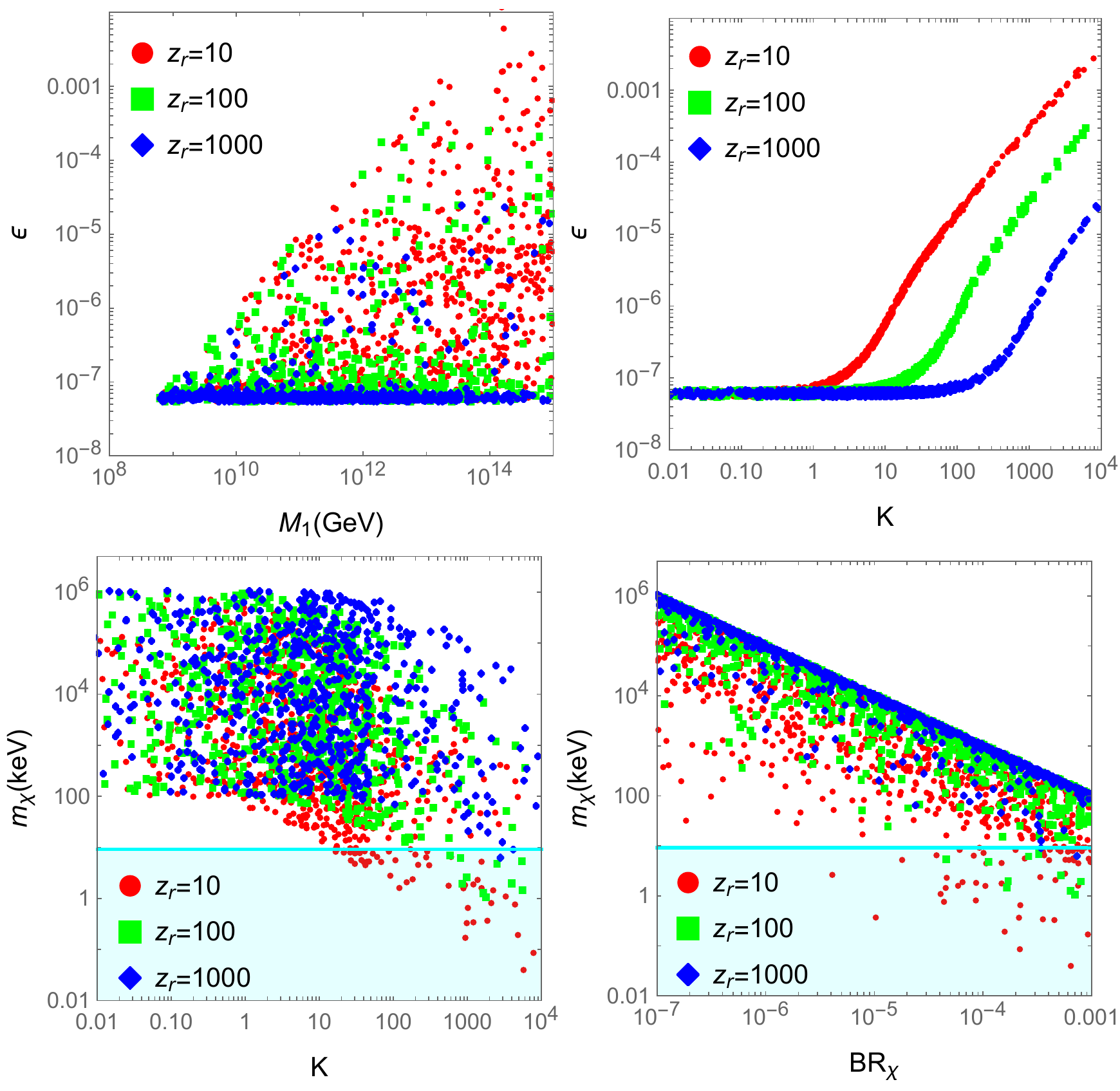}
	\caption{Scanned results for FEU with $z_r=10,100,1000$. Here, we take $n=2$ to illustrate.
		\label{FIG:Scan2}} 	
\end{figure}

The scanned results are illustrated in Fig.~\ref{FIG:Scan1} and Fig.~\ref{FIG:Scan2}, which correspond to the effects of parameter $n$ and $z_r$ respectively.
For the FIMP DM $\chi$, a full analysis of structure formation in Ref.~\cite{DEramo:2020gpr}  set a conservative mass bound $m_\chi>9.2$~keV to satisfy all limits. The exclusion region is marked as the shaded cyan area in Fig.~\ref{FIG:Scan1} and Fig.~\ref{FIG:Scan2}.

The Davidson-Ibarra bound in Eqn.~\eqref{eq:eps1} is clearly shown in the $\epsilon-M_1$ panel. Together with the lower bound $\epsilon \gtrsim 6\times 10^{-8}$ to generate observed $Y_B$, we will have a lower limit on heavy Majorana neutrino mass $M_1\gtrsim 10^9$ GeV. First, let's consider the effects of parameter $n$ as shown in Fig.~\ref{FIG:Scan1}. We observe that the maximum viable CP asymmetry $\epsilon$ in FEU is smaller than in SC. To be concrete, $\epsilon\lesssim3\times10^{-3}$ for SC, $\epsilon\lesssim5\times10^{-4}$ for $n=2$ and $\epsilon\lesssim2\times10^{-5}$ for $n=4$. Due to the FEU, the final lepton asymmetry $Y_L$ is enhanced for the strong washout scenario. Therefore, the required CP asymmetry $\epsilon$ in FEU will become much smaller than in SC for the same value of parameter $K$. The upper limits of $\epsilon$  correspond to the largest viable $K\sim10^4$. In this way, we have a good chance to distinguish different cosmology scenarios. As for DM, a large portion of parameter space with $K\gtrsim 10$ and $m_\chi\lesssim 9.2$ keV in SC is excluded. Meanwhile, the exclusion region in FEU with $n=2$ is much smaller, where the excluded area needs $K\gtrsim100$. And for FEU with $n=4$, the DM bound can hardly exclude the allowed parameter space. This is because a FEU will lead to a suppression of final DM abundance $Y_\chi(\infty)$ and thus an enhancement of DM mass $m_\chi$ when requiring correct relic density as $\Omega_\chi h^2\propto Y_\chi(\infty)m_\chi$. From the distribution of samples in the $m_\chi-\text{BR}_\chi$ panel, a good separation of different cosmology scenario is also possible. In SC, the DM branching ratio BR$_\chi$ as small as $10^{-6}$ can be excluded by the $m_\chi\lesssim 9.2$ keV constraint. Meanwhile, for FEU with $n=2$, the DM constraint can only exclude certain regions with BR$_\chi\gtrsim10^{-5}$. Finally, from the effects of parameter $z_r$ in Fig.~\ref{FIG:Scan2}, we observe similar distributions of viable samples as varying parameter $n$.

\section{Conclusion}\label{Sec:CL}

In this paper, we have systematically investigated the phenomenology of FIMP DM from leptogenesis in the fast expanding Universe. This model is extended by three heavy Majorana neutrinos $N$, which generate the tiny neutrino masses via the type-I seesaw mechanism. The thermal leptogenesis is responsible for generating baryon asymmetry in the Universe. These heavy neutrinos $N$ also mediate the interactions between SM particles and the dark sector, which consists of a scalar $\phi$ and a fermion $\chi$. Here, we have assumed the corresponding Yukawa coupling $\lambda \bar{\chi}\phi N$ is tiny, so that $\chi$ acts as the FIMP DM candidate under the $Z_2$ symmetry. The resulting FEU is obtained by introducing an additional species $\varphi$, whose energy density red-shifts as $\rho_{\varphi}\propto a^{-(4+n)}(n>0)$. The corresponding effects of FEU are controlled by two free parameters $n$ and $z_r$.

In FEU, the dynamics of lepton asymmetry $Y_L$ and DM abundance $Y_\chi$ can be greatly changed. The effects of FEU depend heavily on the decay parameter $K$. For weak washout scenario $K\lesssim1$, modifications of final lepton asymmetry $Y_L$ and DM abundance $Y_\chi$ are relatively small. But for strong washout case $K\gtrsim1$, the final $Y_L(Y_\chi)$ could be increased (suppressed) by several orders of magnitudes. We also introduce the scale factors $F_L$ and $F_\chi$ to describe the corresponding changes. Interestingly, a maximum value of $F_L$ is observed for certain $K$. Using the scale factors, we can easily quantify the washout effects by employing the efficiency factor $	\eta_L^\text{FEU}=\eta_L^\text{SC} F_L^\text{FEU},~\eta_\chi^\text{FEU}=\eta_\chi^\text{SC} F_\chi^\text{FEU}$, where $\eta_L^\text{SC}$ and $\eta_\chi^\text{SC}$ have approximate expressions in Eqn.~\eqref{eq:EF1}.

Explorations of the viable parameter space for success leptogenesis and correct DM relic density in SC and FEU scenarios are also performed in this paper. In FEU, the allowed parameter space is usually smaller than in SC. For example, the maximum value of $\epsilon$ in FEU with $n=2,z_r=100$ is $\sim 5\times 10^{-4}$, about one-tenth of the SC. The best distribution to distinguish different scenarios is considering the $\epsilon-K$ relations from leptogenesis. As for DM, FEU usually leads to heavier $m_\chi$ than in SC, thus can avoid current DM bound as $m_\chi\leq 9.2$ keV more easily. A good separation for DM in the strong washout region is also possible for different cosmology scenarios. On the other hand, the effects of increasing $z_r$ will lead to similar results for enlarging $n$. In short, the impact of FEU on leptogenesis and DM is observable for a strong washout scenario. However, it seems hard to figure out explicit values of the two parameters $n$ and $z_r$ when only considering the results of FIMP DM from leptogenesis.

\section*{Acknowledgments}
This work is supported by the National Natural Science Foundation of China under Grant No. 11805081, Natural Science Foundation of Shandong Province under Grant No. ZR2019QA021 and ZR2018MA047.


\end{document}